\documentclass{appolb}
\usepackage{graphicx}
\usepackage{hyperref}

\begin{document}
\title{Looking forward: Photon induced processes with tagged protons at the CMS experiment
\thanks{Presented at ``Diffraction and Low-$x$", Corigliano Calabro, Italy, 24-30 Sept. 2022.}
}
\author{Michele Gallinaro\\
on behalf of the CMS and TOTEM Collaborations\\
\address{Laborat\'orio de Instrumenta\c{c}\~ao e F\'isica Experimental de Part\'iculas\\
        LIP Lisbon, Portugal}
}
\maketitle
\begin{abstract}
Photon induced processes can be used as a sensitive probe of new physics searches and can be studied using exclusive processes. These processes lead to unprecedented sensitivities on quartic anomalous couplings between photons and W and Z bosons, and new physics searches. By tagging the leading proton from the hard interaction, the Precision Proton Spectrometer (PPS) provides an increased sensitivity to select exclusive processes. PPS is designed to operate in standard high-luminosity runs at the LHC  to perform measurements of e.g. the quartic gauge couplings and search for rare exclusive processes. The first results obtained with PPS, and the status of the ongoing program are discussed.
\end{abstract}
  
\section{Introduction}
Despite its remarkable success, the standard model (SM) is an incomplete theory of Nature and leaves several open questions, such as the nature of Dark Matter, the stability of the Higgs field, the origin of the hierarchy of particle masses, as well as the asymmetry of matter and anti-matter, among them. These facts point at the presence of new physics beyond the SM at an energy scale that could be relatively close to the electroweak scale.
Experiments at the Large Hadron Collider (LHC) have been taking data and produced a wealth of physics results, continuing to perform new measurements and refine searches for new phenomena. A few of the measurements indicate tensions with SM expectations but there is not yet a clear indication for the presence of new physics in the data.
New physics may manifest itself either in direct searches for new phenomena or indirect searches through precise measurements.
A sector of the SM that can be probed at the LHC is that of gauge interactions. In particular, processes involving interactions between photons, Z and W bosons are becoming accessible with the increased amount of data collected at the LHC.
Among those, photon induced processes can be studied at the LHC by tagging the leading proton from the hard interaction.

\section{Central Exclusive Processes and the Precision Proton Spectrometer}
Studies of central exclusive production (CEP) processes in high-energy proton-proton collisions provide a unique method to access a class of physics processes, such as new physics via anomalous production of fermions, V (V=$\gamma$,W,Z) bosons, high transverse momentum ($p_T$) jet production, and possibly the production of new resonances or pair production of possible new particles.

The addition of new detectors further extends the coverage and enhances the sensitivity of the LHC experiments thus offering a new opportunity to explore processes and final states previously not covered.
The Precision Proton Spectrometer (PPS)~\cite{ctppstdr} allows to measure the surviving scattered protons during standard running conditions in regular "high-luminosity" fills. 
It adds precision tracking and timing detectors in the very forward region on both sides of the CMS detector at about 210 meters from the interaction region to study CEP processes in proton-proton collisions.
The PPS detector has been successfully operated since 2016; PPS already collected more than 100~fb$^{-1}$ of data in Run2 in normal high-luminosity proton-proton collisions. 
Thanks to the absence of proton remnants (the proton remains intact), these studies can be performed in clean experimental conditions. 
At the LHC, however, action must be taken to reduce the background 
due to the multiple concurrent soft collisions (pileup, PU) that are superimposed on the hard collision of interest. 
Either additional information on the proton arrival time or kinematics of the final state objects must be used to further suppress the background due to concurrent PU interactions.

\begin{figure*}[htb]
\centerline{
\includegraphics[width=12.5cm]{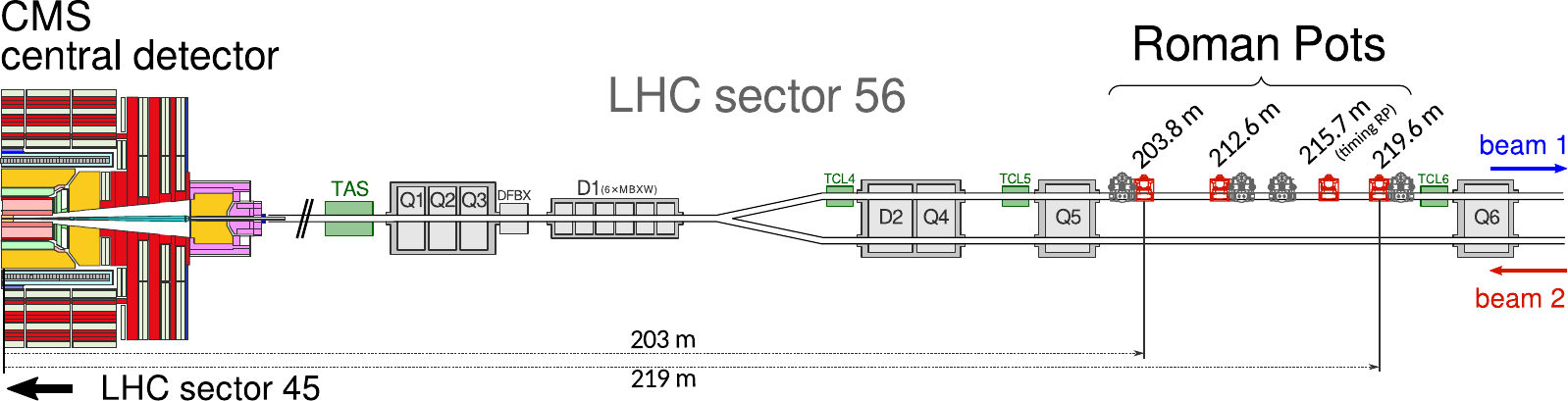}}
\caption{
Schematic layout of the beam line between the interaction point (IP) and PPS ("Roman Pots" in the figure). Although the PPS detector location is symmetric on both sides of the IP, only the negative $z$ direction is shown here.
}
\label{fig:pps}
\end{figure*}

CEP processes are characterized by the presence of leading protons.
By reconstructing the proton(s) kinematics it is possible to correlate the protons with the events in the central CMS detector.
CEP of an object X may occur in the process $pp\rightarrow p + X + p$, where `+" indicates the ``rapidity gaps" adjacent to the state $X$. Rapidity gaps are regions without primary particle production.
In the high-mass region with both protons detected, among some of the most relevant final states are $X = e^+e^-,\mu^+\mu^-,\tau^+\tau^-$, and VV (i.e. $W^+W^- , ZZ, \gamma\gamma$).
In CEP processes, the mass of the state $X$ can be reconstructed from the fractional momentum loss $\xi_1$ and $\xi_2$ of the scattered protons by using the expression $M_X=\sqrt{\xi_1\cdot \xi_2\cdot s}$.
At the LHC, the $M_X$ reach is significantly larger than at previous colliders because of the larger $\sqrt{s}$.
The scattered protons can be observed thanks to their momentum loss, due to the horizontal deviation from the beam trajectory. 
The acceptance in $\xi$ depends on the proximity of the proton detectors to the beam and on the distance from the interaction point (IP).
A sketch of one of the arms of CMS, showing the location of the PPS Roman Pots (RPs) is shown in Fig.~\ref{fig:pps}. 
With the PPS setup, protons that lost 3-15\% of their momentum can be measured.
This translates into an acceptance starting at $M_X \simeq 300$~GeV.
The fractional momentum loss $\xi$ of the protons can be obtained from the proton track
positions and angles (details in Ref.~\cite{TOTEM:2022vox}). 

\section{Exclusive Processes}

Particle production with masses at the electroweak scale via photon-photon fusion can be studied using proton-proton collisions at the LHC. In fact, the LHC is also a photon-photon collider with an energy range of $\sqrt{s_{\gamma\gamma}}$ up to approximately 1-2~TeV, a region so far unexplored.

\subsection{Exclusive diphotons}
Four-photon interactions can be probed with an unprecedented precision using proton tagging with PPS, providing a window on physics beyond the SM.
The elastic scattering of two photons $\gamma\gamma\rightarrow\gamma\gamma$ is a process that takes place via virtual one-loop box diagrams containing charged particles, and its direct observation in the laboratory is difficult. It was observed in heavy ion collisions at the LHC~\cite{ATLAS:2019azn}.
It is worth noticing that the SM predicts the photon-photon production happens through a QCD process via gluon-exchange or QED process via photon-exchange. For a diphoton invariant mass $m_{\gamma\gamma}\geq 100$GeV, i.e. the region accessible to PPS, the latter dominates~\cite{Fichet:2014uka}.
Sensitivity to new massive charged particles can be improved by light-by-light scattering measurement. This can be studied in the EFT framework in terms of the sensitivity to four-photon couplings $\zeta_i$. 
Furthermore, charged particles may contribute to the muon anomalous gyromagnetic moment via higher order (two- and three-loops) diagrams.

A search for exclusive two-photon production via photon exchange in proton-proton collisions with intact protons is performed.
Events are selected with a diphoton invariant mass above 350~GeV and with both protons intact in the final state, to reduce backgrounds from strong interactions. The events of interest are those where the invariant mass and rapidity calculated from the momentum losses of the forward-moving protons match the mass and rapidity of the central, two-photon system.
One exclusive diphoton candidate is observed for an expected background of 1.1 events. Limits at 95\% confidence level (CL) are derived on the four-photon anomalous coupling parameters using an effective field theory. 
Cross section limits of $\sigma<0.61$fb for $m_{\gamma\gamma}\geq 350$GeV are set.
Additionally, axion-like particles are excluded in the mass range of 500 to 2000 GeV~\cite{TOTEMCollaboration:2021xam,CMS:2022zfd}.

\subsection{Exclusive $t\bar{t}$}

At the LHC energies, the dominant top quark production mode is via strong interaction processes, resulting in the production of top quark pairs. Top quarks can also be singly produced in electroweak processes with a smaller cross section, about half that of pair production.
The CEP of top quark pairs is predicted to occur at the LHC with a small cross section (approximately 0.3~fb\cite{Luszczak:2018dfi}) and occurs in proton-proton scattering via the exchange of colourless particles such as photons or pomerons. It receives contributions from QED and QCD diagrams.
Thanks to the correlation between the kinematics of the outgoing intact protons and the central system, the full independent reconstruction of the $t\bar{t}$ system is possible. As the process is sensitive to the electroweak top-photon coupling, it can be used to search for new physics in the EFT or anomalous coupling frameworks, and may offer complementary information to processes like $t\bar{t}\gamma$. 

\begin{figure*}[htb]
\centerline{
\includegraphics[width=12.5cm]{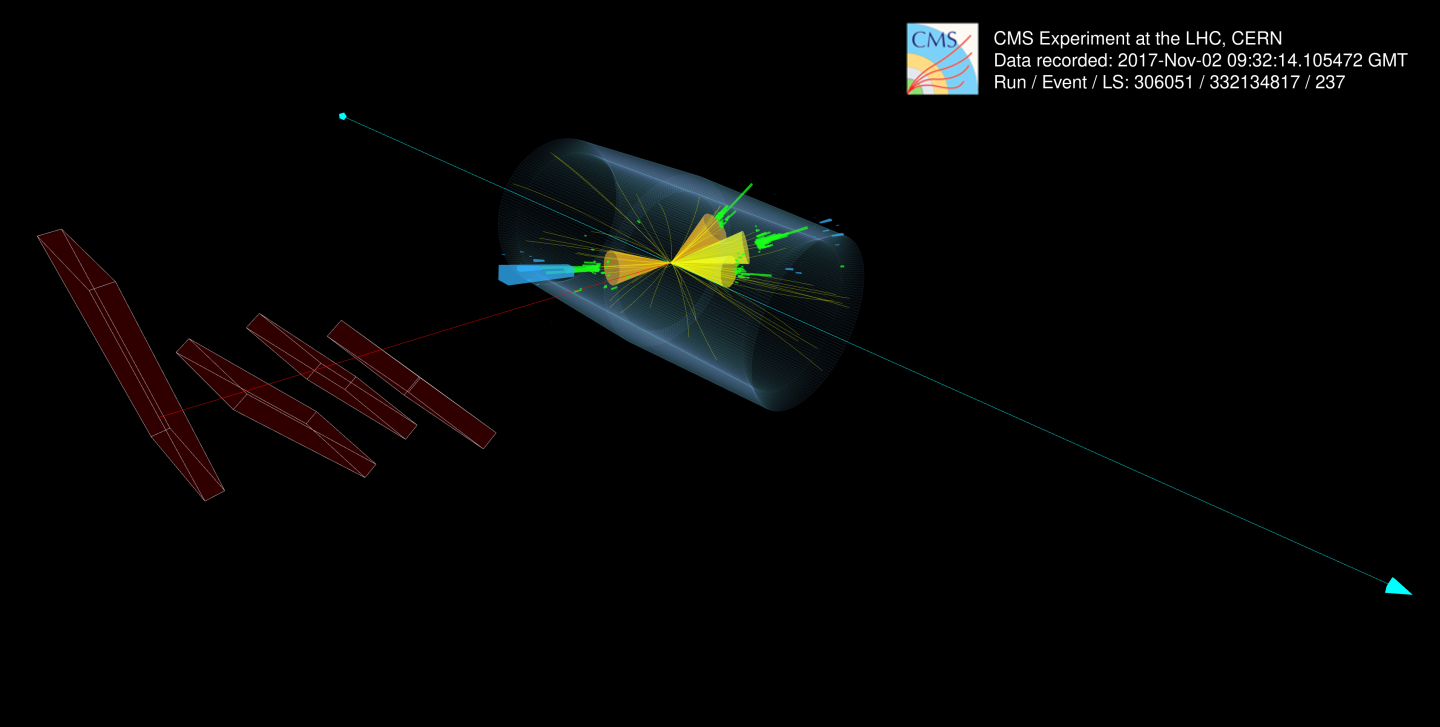}}
\caption{
Event display from a search for $t\bar{t}$ exclusive production in proton-proton collisions at 13 TeV and with tagged protons. The two cyan lines with arrows are the protons that escaped intact from the interaction point and whose tracks have been reconstructed by PPS. The remaining objects are those reconstructed by the central CMS detectors, and constitute the decay products of a $t\bar{t}$ candidate. Top quarks decay almost 100\% of the time to a bottom (b) quark and a W boson. The jets originating from b quarks are shown with the orange cones. One of the two W bosons in this event is reconstructed from its decay in two lighter quarks (up, down, strange, or charm), generating the jets represented by the yellow cones; the other is reconstructed from its decay to a muon, indicated by the red line, and a neutrino, which escapes undetected but causes an imbalance in the visible energy.}
\label{fig:exclusive_top}
\end{figure*}

A search for the central exclusive production of top quark pairs is performed using proton-tagged events in proton-proton collisions at the LHC. The $t\bar{t}$ decay products are reconstructed using the CMS central detector, while forward protons are detected with PPS (Fig.~\ref{fig:exclusive_top}).
At least one of the two W bosons from top quark decays is reconstructed in the $e\nu_e$ or $\mu\nu_\mu$ leptonic decay, while the other is reconstructed either in the leptonic or hadronic decay mode.
The search is conducted separately for the two final states, and the results are afterwards combined. Results are consistent with predictions from the SM, and an upper limit of 0.59 pb at the 95\% confidence level is set on the central exclusive production of $t\bar{t}$ pairs~\cite{2140837}.

\subsection{Exclusive WW and ZZ}

A search is performed for the exclusive production of gauge boson pairs (VV=WW, ZZ) from $\gamma\gamma$ interactions using PPS~\cite{CMS:2022ihy}.
The final state only includes the two bosons and the scattered protons, which are reconstructed using PPS. 
The hadronic decays of the W and Z bosons are studied. When the gauge bosons are produced with a large boost, as expected in many scenarios beyond the SM (BSM), decay products of each of the bosons are merged into a single large-area jet.
The branching fraction of the fully hadronic decays for W and Z bosons is the largest, but this mode is hard to identify without the proton detection because of the very large background from jet production in QCD processes. 
In CEP processes, it is possible to independently reconstruct the kinematics of the final-state bosons both with the central CMS detector and with PPS, when the two protons are reconstructed.

Within the SM, quartic couplings involving two-photon production of charged ($W^\pm$) gauge bosons are allowed at tree level. Because of gauge invariance, the strength of these couplings is related to the triple gauge couplings, and is fully specified in the SM. The SM cross sections for the $\gamma\gamma\rightarrow WW$ and $\gamma\gamma\rightarrow ZZ$ processes at $\sqrt{s}=13$~TeV with both protons intact are expected to be about 50~fb and 50~ab, respectively.
It is worth noticing that SM cross sections are at small masses ($\ll$1 TeV) and therefore the search is not yet really sensitive to exclusive SM VV production despite the 50~fb cross section.
BSM effects on $\gamma\gamma\rightarrow VV$ processes are predicted in a variety of models, with both resonant and nonresonant signals. 
A common approach to quantify deviations from the SM with minimal assumptions involves anomalous quartic gauge couplings (AQGCs).
No significant excess is found over the SM background prediction. The resulting limits are interpreted in terms of dimension-6 and dimension-8 AQGCs. 
The limits are significantly more stringent than those obtained from the $\gamma\gamma\rightarrow WW$ process without proton detection.
For diboson invariant masses $m_{VV}>1$~TeV, cross section limits of $\sigma<67(43)$~fb are set for exclusive WW (ZZ) processes.

\subsection{Exclusive Z/$\gamma+X$}

Searches for new BSM physics processes continue in full force at the LHC. 
In addition to using specific theoretical models, it is also possible to look for new phenomena using model-independent signature-based searches.

A generic search for a hypothetical massive particle $X$ produced in association with one or more SM particles in CEP processes is performed.
In the interaction, the two colliding protons survive after exchanging two colourless particles, and can be recorded in PPS. The detection and precise measurement of both forward protons allow a full kinematic reconstruction of the event, including the four-momentum of $X$ measured from the balance between the tagged SM particle(s) and the forward protons. This technique – the ``missing mass" technique – allows to search for BSM particles without assumptions about their decay properties, except that the decay width can be considered narrow enough to produce a resonant mass peak, thus providing a new tool for generic BSM searches. A search for a massive particle produced in association with a $Z$ or a photon in the final state is considered.
The excellent proton momentum reconstruction of PPS allows to search for missing mass signatures at high invariant masses 
with unprecedented resolution. In this high mass range, electroweak (EWK) processes are generally enhanced relative to QCD-induced processes. The main goal is the search for a photon-photon induced exclusive production process in which an unspecified weakly interacting BSM particle with a narrow decay width is produced. 
No assumption is made on its decay properties.
Leptonically decaying Z bosons or an isolated photon are selected in the central detector, and the missing mass is constructed from the kinematics of the reconstructed boson in the central detector and the final state protons in PPS.
A hypothetical $X$ resonance is searched for in the mass region between 0.6 and 1.6 TeV.
In the absence of significant deviations in data with respect to the background predictions, upper limits on the visible cross section of the $pp \rightarrow pp Z/\gamma + X$ process are set~\cite{CMS-EXO-19-009}. 
For $m_X=600-1600$~GeV, cross section limits of approximately $0.03-0.10 (0.5-1.8)$~pb are set for exclusive ZX ($\gamma$X) processes.

\section{Status and Prospects}

The PPS has been upgraded with new tracking and timing stations and has been taking data in Run3, which started in 2022.
The CMS collaboration intends to pursue the study of CEP events at the High-Luminosity LHC (HL-LHC) by means of a new near-beam proton spectrometer~\cite{CMS:2021ncv}. 
Studies of the main physics topics of interest in the expanded physics program have been performed; R\&D is being pursued to prepare the new spectrometer, explore the feasibility and assess the expected performance. 
Four locations have been identified as suitable for the installation of the new movable proton detectors: at 196, 220, 234, and 420~m from the IP, on both sides. 
Acceptance studies indicate that beams crossing in the vertical plane (instead of horizontal crossing) at
the IP would give access to centrally produced states $X$ in the mass range 43~GeV to 2.7~TeV (or 133~GeV--2.7~TeV if the 420~m station is not included), which makes it possible to study CEP of the 125~GeV Higgs boson. This is a major improvement with respect to the current mass range of 350~GeV to 2~TeV.
Synergies with the other detector upgrade projects will further enhance the physics reach at the HL-LHC.

\section{Summary}

In summary, the LHC is known to collide protons but is also a photon-photon collider.
The study of photon collisions may provide new insights in the nature of the electroweak interactions. 
PPS extends the detector acceptance to very forward regions and provides additional sensitivity to BSM processes.
The detectors are regularly operating in high-luminosity fills and collected approximately 100~fb$^{-1}$ of data in Run~2.
There are challenges on the detector front. The devices themselves have to sustain exceedingly high radiation flux given their proximity to the beam. New detector technology needs to be developed in view of the HL-LHC upgrades. 
The experiment also relies on high-precision timing detectors to accurately identify the primary vertex of the hard collision.
First results of exclusive processes have been obtained and provide for the first time the conditions to study particle production with masses at the electroweak scale via photon-photon fusion.

\section*{Acknowledgements}

To my TOTEM and CMS colleagues who continue to provide continuity and support for the success of this difficult project without forgetting that there are many challenges ahead.
To the Organizers for the kind invitation and an interesting conference in a remote place on this planet.


\begin{thebibliography}{9}

\bibitem{ctppstdr}
CMS and TOTEM Collaborations, ``CMS-TOTEM Precision Proton Spectrometer,''
\href{https://cds.cern.ch/record/1753795}{CERN-LHCC-2014-021}.

\bibitem{TOTEM:2022vox}
 [CMS and TOTEM],
``Proton reconstruction with the CMS-TOTEM Precision Proton Spectrometer,''
\href{https://arxiv.org/abs/arXiv:2210.05854}{arXiv:2210.05854 [hep-ex]}.

\bibitem{ATLAS:2019azn}
G.~Aad et al. [ATLAS],
``Observation of light-by-light scattering in ultraperipheral Pb+Pb collisions with the ATLAS detector,''
Phys. Rev. Lett. 123 (2019) 052001,
\href{https://arxiv.org/abs/arXiv:1904.03536}{arXiv:1904.03536 [hep-ex]}.

\bibitem{Fichet:2014uka}
S.~Fichet, G.~von Gersdorff, B.~Lenzi, C.~Royon and M.~Saimpert,
``Light-by-light scattering with intact protons at the LHC: from Standard Model to New Physics,''
JHEP 02 (2015) 165,
\href{https://arxiv.org/abs/arXiv:1411.6629}{arXiv:1411.6629 [hep-ph]}.

\bibitem{TOTEMCollaboration:2021xam}
A.~Tumasyan et al. [CMS and TOTEM],
``First search for exclusive diphoton production at high mass with tagged protons in proton-proton collisions at $\sqrt{s}=13$~TeV,''
Phys. Rev. Lett. 129 (2022) 011801,
\href{https://arxiv.org/abs/arXiv:2110.05916}{arXiv:2110.05916 [hep-ex]}.

\bibitem{CMS:2022zfd}
 [CMS and TOTEM],
``Search for high-mass exclusive diphoton production with tagged protons,''
\href{http://cds.cern.ch/record/2810862}{CMS-PAS-EXO-21-007,TOTEM NOTE 2022-005}.

\bibitem{Luszczak:2018dfi}
M.~\L{}uszczak, L.~Forthomme, W.~Sch\"afer and A.~Szczurek,
``Production of $ t\overline{t} $ pairs via $\gamma\gamma$ fusion with photon transverse momenta and proton dissociation,''
JHEP 02 (2019) 100,
\href{https://arxiv.org/abs/arXiv:1810.12432}{arXiv:1810.12432 [hep-ph]}.

\bibitem{2140837}
A.~Tumasyan [CMS and TOTEM],
``Search for central exclusive production of top quark pairs in proton-proton collisions at $\sqrt{s}=13$~TeV with tagged protons,'' 
\href{http://cds.cern.ch/record/2803843}{CMS-PAS-TOP-21-007; TOTEM-NOTE-2022-002}.

\bibitem{CMS:2022ihy}
[CMS and TOTEM],
``Search for exclusive $\gamma\gamma \rightarrow WW$ and $\gamma\gamma \rightarrow ZZ$ production in final states with jets and forward protons,''
\href{https://arxiv.org/abs/arXiv:2211.16320}{arXiv:2211.16320 [hep-ph]}.

\bibitem{CMS-EXO-19-009}
[CMS and TOTEM],
``A search for new physics in central exclusive production using the missing mass technique with the CMS-TOTEM precision proton spectrometer,''
\href{http://cds.cern.ch/record/2803840}{CMS-PAS-EXO-19-009, TOTEM NOTE 2022-003}.

\bibitem{CMS:2021ncv}
 [CMS],
``The CMS Precision Proton Spectrometer at the HL-LHC -- Expression of Interest,''
\href{https://arxiv.org/abs/2103.02752}{arXiv:2103.02752 [physics.ins-det]}.

\end{thebibliography}
\end{document}